\documentclass[prl,twocolumn,showpacs,amsmath,amssymb,superscriptaddress]{revtex4-1}
\usepackage{graphicx}
\usepackage{hyperref}
\usepackage{xcolor}
\usepackage{color}
\usepackage[all,cmtip]{xy}
\usepackage{amsthm} 
\usepackage{amsmath}   % <-- for \eqref
\usepackage{amsfonts} 
\usepackage{multirow}
\makeatother

\begin{document}

\title{Quasi-nodal lines in rhombohedral magnetic materials}
\author{Rafael Gonz\'{a}lez-Hern\'{a}ndez}
\email{rhernandezj@uninorte.edu.co}
\affiliation{Departamento de F\'{i}sica y Geociencias, Universidad del Norte, Km. 5 V\'{i}a Antigua Puerto Colombia, Barranquilla 080020, Colombia}
\affiliation{Institut f\"ur Physik, Johannes Gutenberg Universit\"at Mainz, D-55099 Mainz, Germany}
\author{Erick Tuiran}
\email{etuiran@uninorte.edu.co}
\affiliation{Departamento de F\'{i}sica y Geociencias, Universidad del Norte, Km. 5 V\'{i}a Antigua Puerto Colombia, Barranquilla 080020, Colombia}
\author{Bernardo Uribe}
\email{bjongbloed@uninorte.edu.co}
\email{uribe@mpim-bonn.mpg.de}
\affiliation{Departamento de Matem\'{a}ticas y Estad\'{i}stica, Universidad del Norte, Km. 5 V\'{i}a Antigua Puerto Colombia, Barranquilla 080020, Colombia}
\affiliation{Max Planck Institut f\"ur Mathematik, Vivatsgasse 7, 53115 Bonn, Germany}
	
\date{\today}
	
\begin{abstract}
A well-established result in condensed matter physics states that materials crystallizing in symmetry
groups containing glide reflection symmetries possess nodal lines on the energy bands.
These nodal lines are topologically protected and appear on the fixed planes of the reflection in reciprocal space.
In the presence of inversion symmetry, the energy bands are degenerate and the nodal lines
on the fixed plane may hybridize or may cross. In the former case, the crossing is avoided, thus producing
lines on reciprocal space where the energy gap is small, and in the latter, the nodal lines will endure, thus producing
Dirac or double nodal lines. In addition, if the material crystallizes in a ferromagnetic phase where the glide
reflection symmetry is broken, the nodal lines hybridize, thus defining lines in reciprocal space where the energy gap is small.
In this work we concentrate our efforts on the study of nodal lines that hybridize due to magnetization;
we have coined the term of quasi-nodal lines for those lines in reciprocal space where the energy gap is small (less than what
can be detected experimentally). We study magnetic trifluorides and trioxides which crystallize in magnetic space groups 167.107 and 161.71 and we show the existence of quasi-nodal lines on these materials. We furthermore show that whenever the quasi-nodal lines are located around the Fermi level then interesting charge and spin transport effects are induced and can be used to detect experimentally these lines. 
Of particular interest are the half-metallic ferromagnetic phases of PdF$_3$ and LiCuF$_3$ where the large signal of the anomalous Hall conductance 
is due to the presence of the quasi-nodal lines on the Fermi level.
		
\end{abstract}
\maketitle
	
\section{Introduction}

Topological non-trivial states of matter have been investigated intensively in recent years in condensed matter physics
 and materials science \cite{topological-effects,Colloquium-topological-insulators,RevModPhys.88.021004}. The field has grown dramatically after the discovery of topological insulators and it is in continuous
 development with the prediction of diverse topological semimetals phases \cite{PhysRevB.90.205136,Burkov2016}. Topological semimetals are materials with
 gapless bulk states and can be classified in Dirac, Weyl and nodal-line semimetals \cite{Topological-Semimetals,Topological-Materials, semimetals-from-dft}. In contrast to Dirac and Weyl
 semimetals, which have zero-dimensional band crossings, nodal line semimetals have prolonged band crossings along
 unique lines in reciprocal space \cite{Weyl-and-dirac-semimetals}.  In particular, the nodal lines can cross the Brillouin zone in the shape of a closed 
line or a ring \cite{PhysRevLett.120.106403,chains}. These special band crossings can induce exotic phenomena and effects such as ultrahigh mobilities, 
extremely high conductivity, large magnetoresistance and unusual anomalous and spin Hall effects \cite{PhysRevB.96.161105, PhysRevB.96.045127,PhysRevB.103.165104, Burkov2018, PhysRevB.102.035164, Shao2020,PhysRevB.100.125112}.
\\
It has been found that the manifestation of nodal-lines in magnetic and nonmagnetic materials is directly related to the presence, absence or combination of symmetries as time-reversal, inversion, mirror, rotation and partial translation \cite{Fang2015,PhysRevB.99.121106,PhysRevB.95.245208,RevModPhys.88.035005,PhysRevB.97.115125}. Depending on particular planes or lines in the reciprocal space that are protected by a set of these  symmetries, nodal lines can also be categorized into Dirac and Weyl-type \cite{Symmetrydemandedtopologicalnodal-linematerials,Zou2019,RevModPhys.93.025002}. In Dirac nodal lines both inversion symmetry and time-reversal symmetry should be present to guarantee a four degeneration along the band crossing \cite{PhysRevLett.115.036806,PhysRevLett.115.036807,PhysRevLett.117.016602,PhysRevLett.117.096401}. On the other side, in Weyl nodal lines the lack of either time-reversal or inversion symmetry permits the band to split and the
 degeneracy is of degree two \cite{Symmetrydemandedtopologicalnodal-linematerials}. A recent discovery of protected Weyl nodal lines in ferromagnetic materials has attracted great attention because of their potential application in novel spintronic devices \cite{PhysRevLett.123.116401,weyl-line,PhysRevB.103.195115,YANG202143,Belopolski1278}.  Hence, the study of crystalline symmetries that topologically predict nodal lines close to the Fermi level is one of the most important goals in this field.
%
%However, predicted Weyl nodal lines are only symmetry protected only when the spin-orbit coupling is negligible. ???
\\
On the other hand, it is well known that valence and conduction bands with the same symmetry eigenvalues can hybridize,
 leading to anticrossing points in band structures  \cite{PhysRevB.96.235145,Mera2021}. However, depending of the
 system symmetries, these anticrossing points can be extended along the Brillouin zone showing a pattern similar to a nodal line;
 therefore they have been coined quasi-nodal line or nodal-line band anticrossings \cite{anticrossing-line}.
 These quasi-nodal lines could also induce novel electronic and spin transport phenomena as it has been observed in nodal line semimetals \cite{anticrossing-line,PhysRevB.101.075125}. 
 In this manuscript, 
 We predict theoretically and detect computationally the formation of quasi-nodal lines in rhombohedral crystal structures, in particular, IrF$_3$, LaAgO$_3$, PdF$_3$ and LiCuF$_3$, and we foresee that this study can be extended to a 
large set of magnetic and nonmagnetic rhombohedral materials based on \#161 and \#167 space groups such as XF$_3$,
 ABF$_3$, XO$_3$, ABO$_3$ and AB(PO$_4$)$_3$. In addition, using the linear response formalism \cite{Nagaosa2010,Sinova2015}
 we find that the quasi-nodal lines are responsible for a large signal of anomalous and spin Hall conductivity in these
 materials. It is expected that this novel topological transport behavior could lead to the design of novel spintronic devices.
\\

\section{Nodal and quasi-nodal lines on Glide reflection invariant planes}

Nodal lines appear on the fixed planes of glide reflection symmetries in reciprocal space \cite{PhysRevMaterials.3.124204}. They
may hybridize or cross in the presence of inversion symmetry, and they hybridize in the ferromagnetic
phases on which the glide reflection symmetry is broken. In this section we make a summary of the different
types of nodal and quasi-nodal lines induced by glide reflection symmetries.

\subsection{Nodal lines}

Consider a system with spin-orbit coupling (SOC) which is invariant under a glide mirror reflection on a plane $\mathcal{G}$. We can write 
\begin{align} \label{definition mirror reflection}
\mathcal{G}(\mathbf{x}) = \sigma_{\mathbf{n}}(\mathbf{x}) + \mathbf{b},
\end{align}
as the composition of the mirror reflection $\sigma_{\mathbf{n}}$ along the plane perpendicular to the unit normal vector $\mathbf{n}$ and the partial translation by $\mathbf{b}$. Since $\mathcal{G}$ is a glide reflection on the crystal,
we have that $\mathcal{G}^2$ is a translation by the Bravais vector $\sigma_{\mathbf{n}}(\mathbf{b}) + \mathbf{b}$. Let us split the vector $\mathbf{b}$ on its components parallel and perpendicular to $\mathbf{n}$:
\begin{align}
\mathbf{b} = \mathbf{b}_{\mathbf{n}} + \mathbf{b}_{\mathbf{n}^\perp}
\end{align}
and note that $ 2 \mathbf{b}_{\mathbf{n}^\perp} = \sigma_{\mathbf{n}}(\mathbf{b}) + \mathbf{b}$ and therefore $\mathbf{b}_{\mathbf{n}^\perp}$ is half a Bravais lattice vector. Hence we have in momentum coordinates
\begin{align}
\mathcal{G}^2 =-e^{-i 2 \mathbf{k} \cdot  \mathbf{b}_{\mathbf{n}^\perp}}.
\end{align}

Let us consider two consecutive energy bands and let us restrict them to the invariant planes of the
operator $\mathcal{G}$. On these planes we have $\mathcal{G}(\mathbf{k})=  \sigma_{\mathbf{n}}(\mathbf{k})= \mathbf{k}$. 
The band electron energies restricted to these planes are both 2-dimensional and therefore
they intersect generically on a 1-dimensional manifold (the intersection might be empty). Whenever
they intersect we obtain the so-called nodal lines. But how are these nodal lines protected?

Whenever the system preserves the time reversal symmetry $\mathbb{T}$ we may focus our attention
on the time reversal invariant points in momentum space (TRIMs) located on the fixed plane by $\mathcal{G}$.  Note that $\mathcal{G}$ permutes the TRIMs, leaving always at least four fixed. The argument is
the following.  If all TRIMs are fixed by $\mathcal{G}$, then $\mathbf{n}$ is one of the following
three unit vectors:
\begin{align}
(1,0,0), (0,1,0), (0,0,1) \label{n =(1,0,0)}
\end{align}
Now, if $\Gamma$ and $\Gamma'$ are different TRIMs with $\Gamma'=  \sigma_{\mathbf{n}}(\Gamma)$,
then $\mathbf{n}$ is  parallel to $\Gamma-\Gamma'$. Checking the possibilities of $ \sigma_{\mathbf{n}}(\pi,0,0)$ we see that the only possible unit vectors for $\mathbf{n}$ are then the following:  
\begin{align}
{\scriptstyle \frac{1}{\sqrt{2}}}(1,\pm 1, 0), {\scriptstyle \frac{1}{\sqrt{2}}}(0,1,\pm 1),{\scriptstyle \frac{1}{\sqrt{2}}} (\pm 1, 0 ,1)  \label{n =(1,1,0)}
\end{align}
Now, it is important to notice that the reflection $ \sigma_{\mathbf{n}}$ fixes two disconnected planes in momentum space whenever
${\mathbf{n}}$ is one of the vectors presented in \eqref{n =(1,0,0)}, i.e. $k_l=0$ and $k_l=\pi$ for $l \in \{x,y,z\}$, meanwhile it fixes only one connected plane whenever
is one of the vectors presented in \eqref{n =(1,1,0)}, i.e. $k_l\pm k_m =0$ for $l \neq m $ and both in  $ \{x,y,z\}$.

The existence of a nodal line along the fixed plane by $\mathcal{G}$ can be predicted if there are two different TRIMs 
$\Gamma$ and $\Gamma'$ along the 
plane with the property that their $\mathcal{G}$ eigenvalues are different.
If this is the case, the energy band diagrams  of any path joining $\Gamma$ and $\Gamma'$ 
will produce an hourglass, and hence  a band intersection along the path. Now since this argument
works for any path between $\Gamma$ and $\Gamma'$, then a nodal line must exist.
The only requirement for this to happen is that $\mathbf{b}_{\mathbf{n}^\perp} \neq 0$, namely that
the mirror reflection has a glide. If $\Gamma$ is fixed by $\mathcal{G}$, then there must exist
$\Gamma'$ also fixed by $\mathcal{G}$ such that $(\Gamma'-\Gamma) \cdot \mathbf{b}_{\mathbf{n}^\perp} \neq 0$ (this because $\mathcal{G}$ fixes four TRIMs spanning the plane and 
$ \mathbf{b}_{\mathbf{n}^\perp}$ belongs to the plane) and such that 
$(\Gamma'-\Gamma) \cdot 2\mathbf{b}_{\mathbf{n}^\perp} \equiv  \pi \ \mbox{mod} \ 2\pi$; this because
 $\mathbf{b}_{\mathbf{n}^\perp}$ is half Bravais vector and $2(\Gamma'-\Gamma)$ is a reciprocal
lattice vector.

\begin{figure}
\includegraphics[width=8cm]{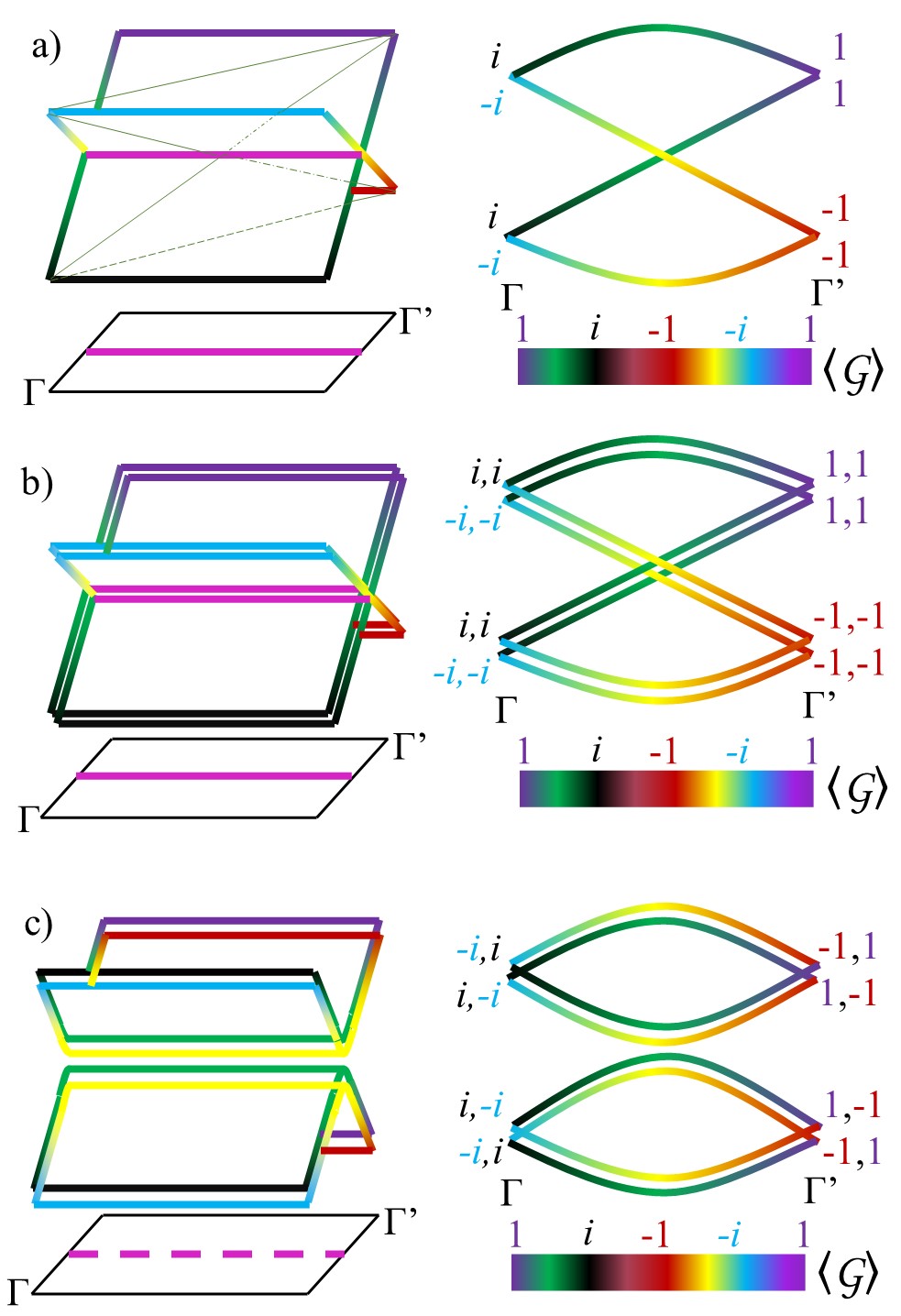}
\caption{Sketch of nodal lines and quasi-nodal lines on glide reflection symmetry planes. (a) The hourglass defined
by the energy bands on any path between
the TRIMs $\Gamma$ and $\Gamma'$ forces the existence of a nodal line on the symmetry plane. In the presence of inversion
the energy bands degenerate and the two bands have the same $\mathcal{G}$ eigenvalue (b) or eigenvalues of
opposite sign (c). Whenever the degenerate bands have same $\mathcal{G}$ eigenvalue, then the existence of a double
nodal line is enforced (b). Whenever the degenerate bands have $\mathcal{G}$ eigenvalues of opposite sign,
the bands hybridize and the crossing is avoided (c). In the latter case, whenever the energy gap is small
($\sim 25$ meV), we call the anticrossings {\it quasi-nodal lines}. In all three panels the colored bar
parametrizes the $\mathcal{G}$ eigenvalues of the energy bands. In panels b) and c) the bands are degenerate but
for the sketch's clarity the bands have been separated.
} 
\label{sketch of nodal lines}
\end{figure}

On $\Gamma$ and $\Gamma'$ the time reversal operator $\mathbb{T}$ conjugates the eigenvalues of $\mathcal{G}$,
and since the eigenvalues  of $\mathcal{G}$ differ by a sign, the only options for the eigenvalues of $\mathcal{G}$
at $\Gamma$ and $\Gamma'$ are $\{1,-1\}$ and $\{i,-i\}$. Hence any path from $\Gamma$ to $\Gamma'$ on
the fixed plane by $\mathcal{G}$ induces an hourglass combinatorial diagram on the energy bands.
The intersection of the middle bands is enforced because the eigenvalues of $\mathcal{G}$ at that point differ by a sign,
and therefore hybridization (or repulsion) is avoided \cite{Gonzalez-Tuiran-Uribe-2021}.  A schematic diagram
of the nodal line thus formed is presented in Fig. \ref{sketch of nodal lines} (a).

Let us see some explicit examples:

a) Consider $\mathcal{G}(x,y,z)=(-x, y, z +{\scriptstyle \frac{1}{2}})$ with $\mathbf{n}=(1,0,0)$ and $\mathbf{b}=
\mathbf{b}_{\mathbf{n}^\perp} = (0,0,{\scriptstyle \frac{1}{2}})$. We have $\mathcal{G}^2=-e^{ik_z}$ and
the fixed planes are $k_x=0$ and $k_x=\pi$. In both planes we get nodal lines that must intersect any path
joining TRIMs with $k_z=0$ to TRIMs with $k_z=\pi$; therefore these nodal lines should cross the fixed plane
along the $k_y$-direction.

b) Note that the previous argument can be applied without any change to the operator
$\mathcal{G}(x,y,z)=(-x+{\scriptstyle \frac{1}{2}}, y, z +{\scriptstyle \frac{1}{2}})$
where $\mathbf{b}_{\mathbf{n}} = ({\scriptstyle \frac{1}{2}},0,0)$ and $\mathbf{b}_{\mathbf{n}^\perp} = (0,0,{\scriptstyle \frac{1}{2}})$. The component of the translation along the direction of $\mathbf{n}$ plays no role on the existence
of the nodal lines protected by the hourglasses. This symmetry appears in the space group \#14 $(P2_1/c)$.

c) Consider $\mathcal{G}(x,y,z)=(y+{\scriptstyle \frac{1}{2}}, x+{\scriptstyle \frac{1}{2}}, z +{\scriptstyle \frac{1}{2}})$ with $\mathbf{n}={\scriptstyle \frac{1}{\sqrt{2}}}(1,-1,0)$ and $\mathbf{b}=
\mathbf{b}_{\mathbf{n}^\perp} = ({\scriptstyle \frac{1}{2}},{\scriptstyle \frac{1}{2}},{\scriptstyle \frac{1}{2}})$. We have $\mathcal{G}^2=-e^{-i(k_x+k_y+k_z)}$ and
the fixed plane is $k_x=k_y$. The nodal lines must meet all paths joining TRIMs along $k_z=0$ with TRIMs along $k_z=\pi$

Note that whenever the translation vector $\mathbf{b}$ is parallel to the reflection vector $\mathbf{n}$, the
 $\mathcal{G}$ eigenvalues are constant $\pm i$ along the fixed planes. Therefore if the energy bands with different
eigenvalues intersect, then there cannot be hybridization and these nodal lines that appear are called {\it accidental}. 

In the presence of the inversion symmetry, all the energy bands are degenerate by Kramer's rule. Some
of the nodal lines previously described survive meanwhile in other cases the bands hybridize thus avoiding the nodal line.
In the latter case, we get what is known as {\it anticrossings} and they are of interest whenever the energy band gap is small.

\subsection{Double nodal lines} 

Let us now suppose that we have time reversal symmetry $\mathbb{T}$, inversion symmetry $\mathcal{I}$ and 
the glide reflection $\mathcal{G}$ as in Eqn. \eqref{definition mirror reflection}. In position coordinates we have
\begin{align}
\mathcal{G}( \mathcal{I}(\mathbf{x})) = \mathcal{I}( \mathcal{G}(\mathbf{x})) + 2 \mathbf{b},
\label{commutator S, I position}
\end{align}
implying thus that $\mathbf{b}$ is also half a Bravais lattice vector (hence we have that all three
vectors $\mathbf{b}$, $\mathbf{b}_{\mathbf{n}}$ and $\mathbf{b}_{\mathbf{n}^\perp}$ are half Bravais vectors); therefore in momentum coordinates
\begin{align} \label{commutator S, I momentum}
e^{-i 2 \mathbf{k}\cdot \mathbf{b}} \mathcal{I}\mathcal{G} = \mathcal{G}\mathcal{I}. 
\end{align}

The composition of the inversion with the time reversal operator $\mathbb{T} \mathcal{I}$ leaves momentum coordinates
fixed and squares to $-1$, thus endowing the energy eigenvalues of the Hamiltonian with a quaternionic structure. This implies
that all energy states come in degenerate pairs due to Kramer's rule. Let us see what happens to the $\mathcal{G}$ eigenvalues
of a Kramer pair once restricted to the fixed planes of $\mathcal{G}$ whose equations are
\begin{align} \label{equations planes fixed by S}
2 \mathbf{n} \ (\mathbf{k} \cdot \mathbf{n})  \equiv 0 \ \mbox{mod} \ \mathbf{G}
\end{align}
with $\mathbf{G}$ reciprocal lattice vectors.

From Eqn. \eqref{commutator S, I momentum} we obtain the commutation relation between $\mathcal{G}$ and
$\mathbb{T} \mathcal{I}$:
\begin{align} \label{commutator S, TI}
e^{-i 2 \mathbf{k}\cdot \mathbf{b}} (\mathbb{T}\mathcal{I})\mathcal{G} = \mathcal{G}(\mathbb{T} \mathcal{I}).
\end{align}
Now let us  consider an eigenfunction of the Hamiltonian $\Psi$ and let us restrict it to the fixed point planes
of the operator $\mathcal{G}$ shown in Eqn. \eqref{equations planes fixed by S}.
On these planes we may diagonalize $\Psi$ as follows:
\begin{align} \label{eigenvalue eqn for S, SOC}
\mathcal{G} \Psi (\mathbf{k}) = \pm i e^{-i  \mathbf{k} \cdot  \mathbf{b}_{\mathbf{n}^\perp}}\Psi (\mathbf{k}).
\end{align}
Replacing $\Psi$ in Eqn. \eqref{commutator S, TI} we obtain the $\mathcal{G}$ eigenvalues for $(\mathbb{T} \mathcal{I}) \Psi$ on the fixed planes:
\begin{align}
\mathcal{G} \left( (\mathbb{T} \mathcal{I}) \Psi(\mathbf{k}) \right)  = 
\mp\left( e^{-i2 \mathbf{k} \cdot \mathbf{b}_{\mathbf{n}}}  \right)  i  e^{-i  \mathbf{k} \cdot  \mathbf{b}_{\mathbf{n}^\perp}}\left((\mathbb{T} \mathcal{I})\Psi (\mathbf{k}) \right).
\end{align}

We see that the $\mathcal{G}$-eigenvalues of both $\Psi$ and its Kramer pair $(\mathbb{T} \mathcal{I})\Psi$ differ by the
 phase factor $-e^{-i2 \mathbf{k} \cdot \mathbf{b}_{\mathbf{n}}}$. This phase is always $-1$ except in the case
that $\mathbf{b}_{\mathbf{n}}\neq 0$ ,the unit normal vector $\mathbf{n}$ is in the list of \eqref{n =(1,0,0)} and the fixed plane is $k_l=\pi$; in this case the phase factor is $1$.

Therefore the pair of  bands $\Psi$ and $(\mathbb{T} \mathcal{I})\Psi$ have always opposite sign $\mathcal{G}$ eigenvalues,
except on the case that $\mathbf{b}_{\mathbf{n}}\neq 0$,  $\mathbf{n}$ is in \eqref{n =(1,0,0)} and the fixed plane is $k_l=\pi$. In this case we have that
$2 \mathbf{k} \cdot \mathbf{b}_{\mathbf{n}} = \mathbf{k} \cdot \mathbf{n}= \pi$ and therefore the bands  
$\Psi$ and it $(\mathbb{T} \mathcal{I})\Psi$  have the same $\mathcal{G}$ eigenvalue.

We may therefore infer that unless $\mathbf{b}_{\mathbf{n}}\neq 0$,  $\mathbf{n}$ is in \eqref{n =(1,0,0)} and the fixed plane is $k_l=\pi$, the energy crossing between a pair of double bands along the fixed planes of $\mathcal{G}$ is avoided
due to hybridization. Each double band has both $\mathcal{G}$ eigenvalues, therefore energy repulsion (hybridization) occurs at all points where the double bands get closer in energy (see Fig. \ref{sketch of nodal lines} (c) for a
schematic diagram of this hybridization). This effect is also called {\it anticrossing of bands} since
the energy gap thus formed between the bands may be very small and therefore a nodal line might be computational and 
experimentally detected. The gap thus formed depends on the intensity of the SOC and therefore
it may be small. Whenever the energy gap of these anticrossings is small the physical effects are relevant and
therefore we may call these anticrossings with the name {\it quasi-nodal line}. The rhombohedral trifluorides and trioxides fall into this category of materials.

Whenever $\mathbf{b}_{\mathbf{n}}\neq 0$,  $\mathbf{n}$ is in \eqref{n =(1,0,0)} and the fixed plane is $k_l=\pi$, the
$\mathcal{G}$ eigenvalues of the Kramer degenerate pairs are equal. Therefore if the reflection symmetry $\mathcal{G}$
predicted nodal lines as presented before, inclusion of the inversion operator keeps the nodal lines. Now they
are nodal lines of double degenerate Kramer's pairs. Originally the nodal lines were protected by the hourglass
argument that produces $\mathcal{G}$, and since the Kramer's pairs have the same $\mathcal{G}$ eigenvalues, the
hourglasses are now double degenerate and also protected (see Fig. \ref{sketch of nodal lines} (b) for a schematic
diagram of these double nodal lines).

Let us see what does the inclusion of the inversion symmetry do on the examples we presented on the previous section.
For the reflections considered in a) and c), namely $\mathcal{G}(x,y,z)=(-x, y, z +{\scriptstyle \frac{1}{2}})$ and 
$\mathcal{G}(x,y,z)=(y+{\scriptstyle \frac{1}{2}}, x+{\scriptstyle \frac{1}{2}}, z +{\scriptstyle \frac{1}{2}})$ respectively,
the absence of translation on the normal direction implies that the Kramer pairs of energy bands along the fixed planes
have opposite $\mathcal{G}$ eigenvalues. Therefore the energy crossings induced by the operator $\mathcal{G}$ hybridize
and we obtain nodal line anticrossings as the ones presented in Fig. \ref{sketch of nodal lines} (c). The energy gap on these anticrossings depends on the intensity of the SOC, hence if the energy gap is small, these anticrossing may well behave like nodal lines (quasi-nodal lines). Nevertheless, they will not be
topologically protected.

For the glide reflection considered in b) with $\mathcal{G}(x,y,z)=(-x+{\scriptstyle \frac{1}{2}}, y, z +{\scriptstyle \frac{1}{2}})$
we will have a nodal line anticrossings on the plane $k_x=0$ like the one in Fig. \ref{sketch of nodal lines} (c), meanwhile the double nodal
 lines appear on the plane $k_x=\pi$ as presented in Fig. \ref{sketch of nodal lines} (b).
On $k_x=0$ the Kramer pair of energy bands possesses opposite $\mathcal{G}$ eigenvalues, meanwhile on $k_x=\pi$
the eigenvalues are the same.

\subsection{Quasi-nodal lines}

We have seen in the previous sections how nodal lines appear and are topologically protected whenever
there is a glide reflection symmetry and time reversal symmetry on the system. These
nodal lines posses different properties in the presence of the inversion symmetry. The composition
of the inversion with time reversal induces Kramer's degeneracy on the energy bands localized on the 
fixed planes of the glide reflection and depending on the type of glide reflection, the Kramer's pairs
may have equal eigenvalues for the glide reflection operator or eigenvalues with opposite sign. In the former
case, the double nodal lines that appear are topologically protected and they are also known as 
{\it Dirac nodal lines} \cite{PhysRevLett.115.036806}; in the latter case the bands hybridize and the band crossing is avoided. 

The existence of protected nodal lines and double nodal lines close to the Fermi level induce exotic 
spin and electronic transport properties \cite{PhysRevB.101.075125} on the material, among them resonant spin-flipped reflection \cite{PhysRevLett.121.166802} and anomalously Hall currents \cite{PhysRevB.97.161113}. 
Therefore the existence of protected nodal lines has been extensively studied in the last years \cite{RevModPhys.93.025002, PhysRevMaterials.3.124204}. 
On the other hand, nodal lines that hybridize due to the presence of the inversion symmetry have seldomly been studied \cite{Symmetrydemandedtopologicalnodal-linematerials}. The energy gap that appears due to the hybridization makes them of less interest.
Nevertheless, whenever the gap that is opened due to hybridization is small, interesting electronic properties are induced on the 
material.

We have therefore coined the name of {\it quasi-nodal lines} for the lines on fixed $\mathcal{G}$ planes that exist whenever
there is a hybridization of nodal lines and whose energy gap is very small (comparable
to room temperature $\sim 25$ meV). It could be argued that since these quasi-nodal lines
are not topologically protected, their existence may not have any implication on the electronic properties of the materials.
We would like to argue otherwise: since the energy gap opened due to hybridization is very small, the electronic properties
detected are similar to the ones observed on nodal lines. Furthermore, if the quasi-nodal lines are close to the Fermi level,
interesting phenomena on charge and spin transport are observed such as large anomalous and Nernst Hall effect, spin Hall effect, among others \cite{PhysRevB.101.075125,PhysRevB.99.165117,anticrossing-line}.

Rhombohedral materials crystallizing on symmetry groups \# 161 and \#167 and ferromagnetic groups \#161.71 and \#167.107
show the presence of quasi-nodal lines on their electronic structure. This is the content of the next section.

\section{Quasi-nodal lines on rhombohedral materials}

Quasi-nodal lines are present on rhombohedral materials crystallizing on the symmetry groups \# 161 and \#167 and on their ferromagnetic phases
 \#161.71 and \#167.107. Among the many materials crystallizing on these symmetry groups, the rhombohedral trifluorides and trioxides make a pair of interesting families of compounds to study. In both families there are compounds crystallizing in all four symmetry groups. In Table \ref{table four groups} we have summarized the properties of the symmetry groups and the materials with quasi-nodal lines on glide reflection planes.

The crystal symmetries that generate these four  groups are 
\begin{align}
\mathcal{C}(x,y,z) &= (y,z,x)  \\
\mathcal{I}(x,y,z) &= (-x,-y,-z)  \\
\mathcal{G}(x,y,z) &= (y+{\scriptstyle \frac{1}{2}},x+{\scriptstyle \frac{1}{2}},z+{\scriptstyle \frac{1}{2}}) \\
\mathcal{R}(x,y,z) &= (-y+{\scriptstyle \frac{1}{2}},-x+{\scriptstyle \frac{1}{2}},-z+{\scriptstyle \frac{1}{2}}) 
\end{align}
with $\mathcal{C}$ a 3-fold rotation, $\mathcal{I}$ inversion, $\mathcal{G}$ a glide reflection and $\mathcal{R}= \mathcal{G}\mathcal{I}$. In momentum coordinates we have 
\begin{align}
{\mathcal{R}}(k_x,k_y,k_z) &= (k_y,k_x,k_z) \\
{\mathcal{G}}(k_x,k_y,k_z) &= (-k_y,-k_x,-k_z),
\end{align}
and the non-trivial relations (including SOC) among the generators are the following:
\begin{align}
\mathcal{I}^2=1, \ \ \mathcal{C}^3=& -1, \  \ \mathcal{R}^2=-1,\ \ \ \mathcal{C}\mathcal{G}=\mathcal{G}\mathcal{C}^{-1}\\
\mathcal{C}(\mathbb{T}\mathcal{G})&=(\mathbb{T}\mathcal{G})\mathcal{C}^{-1} \label{TS vs C}\\
\mathcal{G}^2 &=-e^{-i\pi(k_x+k_y+k_z)} \\
(\mathbb{T}\mathcal{G})^2 &=e^{-i\pi(k_x+k_y+k_z)}  \label{TS^2}\\
\mathcal{R}\mathcal{I} &=e^{i\pi(k_x+k_y+k_z)}\mathcal{I}\mathcal{R},\\ 
\mathcal{G}\mathcal{I} &=e^{-i\pi(k_x+k_y+k_z)}\mathcal{I}\mathcal{G},\\
( \mathbb{T}\mathcal{G})\mathcal{I} &=e^{-i\pi(k_x+k_y+k_z)}\mathcal{I}(\mathbb{T}\mathcal{G}). \label{TS vs I}
\end{align}
Here we have $\mathbf{b}_{\mathbf{n}^\perp} = ( {\scriptstyle \frac{1}{2}},{\scriptstyle \frac{1}{2}},{\scriptstyle \frac{1}{2}})$,
 $\mathbf{b}_{\mathbf{n}}=(1,-1,0)$ and the $\mathcal{G}$ fixed planes are $k_x=k_y$, $k_y=k_z$ and $k_z=k_x$.
The quasi-nodal lines appear on the planes fixed by $\mathcal{G}$ and for all the four symmetry groups they have a multiplicity of 6 on the whole Brillouin zone. This is shown in Table \ref{table four groups}.

\begin{table}
\begin{tabular}{||c||c|c|c|c||} % |p{3cm}||p{3cm}|p{3cm}|p{3cm}|  }
\hline
\multicolumn{5}{||c||}{Rhombohedric crystals with quasi-nodal lines}\\
 \hline
   & \multicolumn{2}{|c|}{No Magnetic} & \multicolumn{2}{|c||}{Ferromagnetic} \\
 \hline
 SG \# & 167 & 161 & 167.107 & 161.71\\
\hline
Generators &  $\mathcal{G},\mathcal{I}, \mathbb{T}, \mathcal{C}$ & $\mathcal{G}, \mathbb{T}, \mathcal{C}$ & $\mathbb{T}\mathcal{G},\mathcal{I},\mathcal{C}$ &$ \mathbb{T}\mathcal{G},\mathcal{C}$ \\
\hline
Symmetries & 24 & 12 & 12 &6\\
\hline
Fixing planes & $\mathbb{T}\mathcal{I}$, $\mathcal{G}$ & $\mathcal{G}$ &  $\mathcal{I} \mathbb{T} \mathcal{G}$ & \\
 \hline
Multiplicity of &\multirow{2}{*}{6}&\multirow{2}{*}{6}&\multirow{2}{*}{6}&\multirow{2}{*}{6}\\
quasi-nodal line &&&&\\
\hline
\multirow{5}{*}{Materials} & IrF$_3$ & \textcolor{red}{LaAgO$_3$}& \textcolor{red}{PdF$_3$},\textcolor{red}{MnF$_3$} & \textcolor{red}{LiCuF$_3$}\\
& InF$_3$  &  \textcolor{red}{LaCuO$_3$}& \textcolor{red}{MnBO$_3$},NiF$_3$   & LiVF$_3$\\
&RhF$_3$& NaCdF$_3$  &\textcolor{red}{LaMnO$_3$},VF$_3$    & \\
&GaF$_3$& CaTlF$_3$ &  \textcolor{red}{TiBO$_3$},RuF$_3$&\\
&AlF$_3$& CsPbF$_3$  & \textcolor{red}{LaNiO$_3$},MoF$_3$   &\\
&ScF$_3$&  & CoF$_3$,RuF$_3$  &\\
& &  & FeF$_3$,CrF$_3$  &\\
\hline
\end{tabular}
\caption{Rhombohedric crystals with quasi-nodal lines on the 
planes fixed by the glide reflection $\mathcal{G}$. The highlighted materials in red have nodal lines crossing the Fermi level.
The group with the most symmetries is \#167, which includes the 3-fold rotation $\mathcal{C}$, inversion $\mathcal{I}$,
time reversal $\mathbb{T}$ and the glide reflection $\mathcal{G}$; whenever the inversion symmetry is broken we obtain
the group \# 161. In the presence of magnetization along the axis of rotation of the 3-fold symmetry $\mathcal{C}$,
time reversal $\mathbb{T}$ is broken together with the glide reflection $\mathcal{G}$. The symmetry that
is kept is the composition $\mathbb{T} \mathcal{G}$ thus defining the magnetic symmetry groups \#167.107 
in the presence of inversion, and \#161.71 whenever the inversion is broken. The quasi-nodal lines appear on the planes
fixed by $\mathcal{G}$; their multiplicity is calculated as the number of total symmetries divided by the number
of symmetries that fix the planes. In all four cases the multiplicity is 6.} 
\label{table four groups}
\end{table}

\begin{figure*}
	\includegraphics[width=14cm]{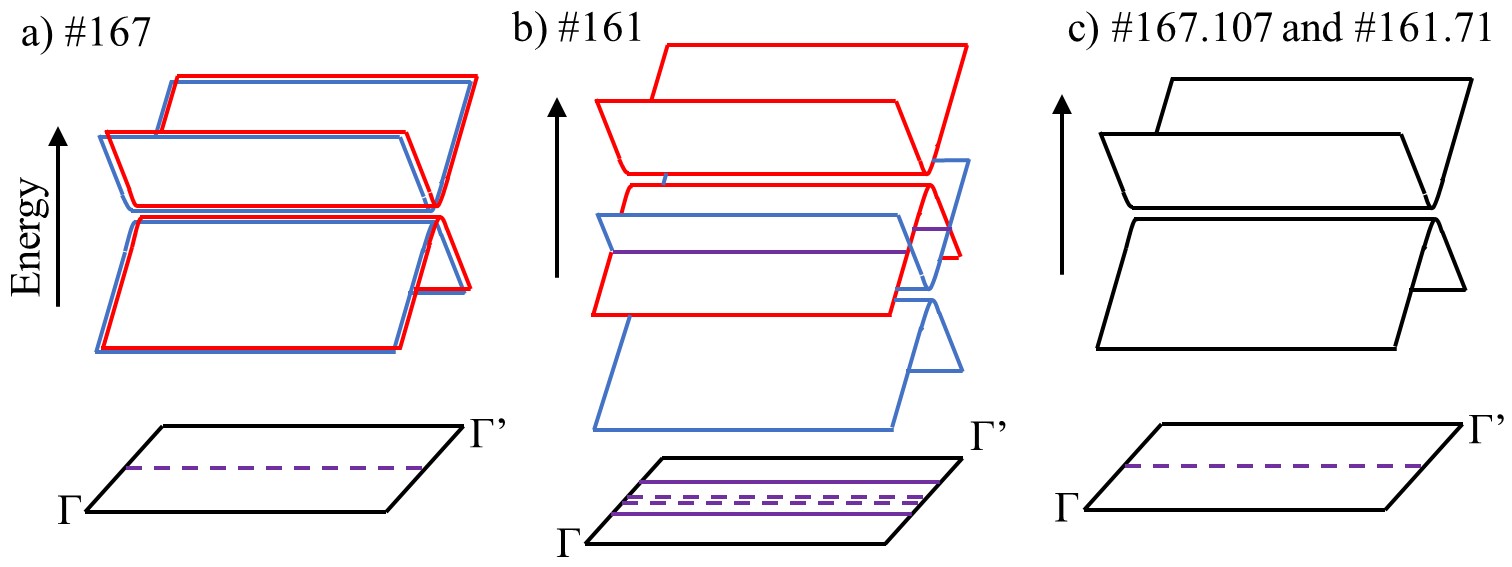}
	\caption{Sketch of quasi-nodal lines on the four symmetry groups 167, 161, 167.107 and 161.71.
		(a) On the SG \#167 the bands are degenerate and the eigenvalues of the operator $\mathcal{G}$
		have opposite sign which permits the bands to hybridize. (b) On the SG \#161 the inversion
		symmetry is broken and the degenerate bands unfold. The bands with same color hybridize and the bands
		with opposite color intersect. Here we observe a pair of accidental nodal lines and a pair of quasi-nodal lines.
		In these two cases the red and blue bands have opposite $\mathcal{G}$ eigenvalue.
		(c) On the magnetic SG \# 167.107 and 161.71 the glide reflection $\mathcal{G}$ symmetry
		is broken and therefore there is no restriction on the bands for their hybridization. In all
		four symmetry groups the quasi-nodal lines appear on the planes fixed by the glide
		reflection $\mathcal{G}$. These quasi-nodal lines have been detected on the non-magnetic
		and ferromagnetic rhombohedral trifluorides presented in Fig. \ref{Fig1}.
	}
	\label{sketch of quasi nodal lines}
\end{figure*}

Due to the presence of the composition of inversion with time reversal on the
materials with SG \#167, the quasi-nodal lines appearing are Dirac quasi-nodal lines or double quasi-nodal lines.
Whenever inversion is broken, as in the materials with SG \#161, the Dirac quasi-nodal lines 
unfold in a pair of quasi-nodal lines and a pair of accidental nodal lines. In Fig. \ref{sketch of quasi nodal lines} a schematic diagram of each of the quasi-nodal lines
on the four symmetry groups has been presented.

In the ferromagnetic rhombohedric trifluorides LiCuF$_3$ and PdF$_3$ (see Fig. \ref{Fig1}) that we present in the next section, the presence
of the quasi-nodal lines has been observed. The valence and conduction bands restricted to the $\mathcal{G}$ fixed plane
can be seen in Fig \ref{Fig2} (a) and (a'), the quasi-nodal lines on the $\mathcal{G}$ plane in Fig. \ref{Fig2} (c) and (c'),
the graph of the nodal line versus the energy in Fig. \ref{Fig2} (f) and (f'),  the hybridization of the energy bands on the anticrossings in Fig. \ref{sketch of quasi nodal lines} (c), \ref{Fig2} (d) and (d'), and 
the six-fold multiplicity of the quasi-nodal lines on the whole
Brillouin zone  in Fig. \ref{Fig2} (e) and (e').

The energy gaps along the quasi-nodal lines presented in Fig. \ref{Fig2} (e) and (e') are less than 0.6 meV in the case of LiCuF$_3$ and
less of 3.0 meV in the case of PdF$_3$. In both cases the quasi-nodal lines include Weyl points lying on the line 
$\Gamma$-T is fixed by the $\mathcal{C}$ rotation. These Weyl points lie close to the point T and their existence can
be deduced from the fact that the $\mathcal{C}$ eigenvalues are different on the bands close to the Fermi level.
Since the $\mathcal{C}$ eigenvalues are different, the hybridization cannot be carried out and the crossing is
topologically protected.
These energy crossings along the symmetry line $\Gamma$-T close
to the point T can be observed at the right hand side of all four panels in Fig. \ref{Fig1}. Here the colors on the energy bands represent different $\mathcal{C}$ eigenvalues.
The fact that the quasi-nodal lines include Weyl points implies that the bandgap energy along the quasi-nodal lines remains small.
The existence of the Weyl points along the line $\Gamma$-T protects the quasi-nodal lines from being completely gapped. 

Let us finish this section with an analysis of the corepresentations that appear on the 
points invariant by the inversion operator, i.e. $\Gamma(0,0,0)$, L$(0,0,\pi)$
F$(0,\pi,\pi)$ and T$(\pi,\pi,\pi)$, and on the lines fixed by $\mathcal{C}$ and $\mathbb{T}\mathcal{G}$. 

The operator $\mathbb{T}\mathcal{G}$ has for fixed points the lines $k_x+k_y=0$ on the planes
$k_z=0$ and $k_z=\pi$. By Eqn. \eqref{TS^2} we see that $(\mathbb{T}\mathcal{G})=1$ on
$\Gamma$ and F and $(\mathbb{T}\mathcal{G})=-1$ on L and T. 
The $k_x=k_y=k_z$ axis is fixed by the rotation operator $\mathcal{C}$ and the eigenvalues of $\mathcal{C}$ are $e^{\frac{i l \pi}{3}}$ with $l=1,3,5$. From the commutation relations presented in Eqn. \eqref{TS vs C} and \eqref{TS vs I} we deduce
that on the points $\Gamma$ and F the antiunitary operator $\mathbb{T}\mathcal{G}$ behaves like the
complex conjugation operator $\mathbb{K}$, the inversion operator like multiplication by $\pm1$, and on $\Gamma$, the operator
$\mathcal{C}$ is multiplication by $e^{\frac{i l \pi}{3}}$ with $l=1,3,5$. Hence on $\Gamma$ and F
the corepresentations are one-dimensional.

On T and L Eqn. \eqref{TS vs I} implies that $\mathbb{T}\mathcal{G}$ changes the sign of the $\mathcal{I}$ 
eigenvalue and therefore the corepresentations are two-dimensional with the following matrix representation:
\begin{align}
\mathcal{I} &= \left(\begin{matrix} +1 & 0 \\ 0 & -1 \end{matrix} \right), & 
\mathbb{T}\mathcal{G} &= \left(\begin{matrix} 0 & 1 \\ -1 & 0 \end{matrix} \right) \mathbb{K}.
\end{align}
On T, because of Eqn. \eqref{TS vs C}, the $\mathcal{C}$ eigenvalues are repeated and its matrix representation is:
\begin{align}
\mathcal{C} &= \left(\begin{matrix} e^{\frac{i l \pi}{3}} & 0 \\ 0 & e^{\frac{i l \pi}{3}} \end{matrix} \right).
\end{align}

On the right hand side of all four panels of Fig. \ref{Fig1} it can be observed that the energy bands with same $\mathcal{C}$ eigenvalue
join at T, thus agreeing with the previous matrix description, and that the energy crossings close to T with different $\mathcal{C}$ eigenvalues define Weyl points \cite{Gonzalez-Tuiran-Uribe-2021}

Let us finally note that whenever we have the presence of time reversal and inversion, all bands are degenerate and the corepresentations
defined above will appear with their Kramer's dual. On T and L points for the symmetry group \#167 the energy bands
have degeneracy four.

\section{Materials realization}

\begin{figure*}
	\includegraphics[width=18.4cm]{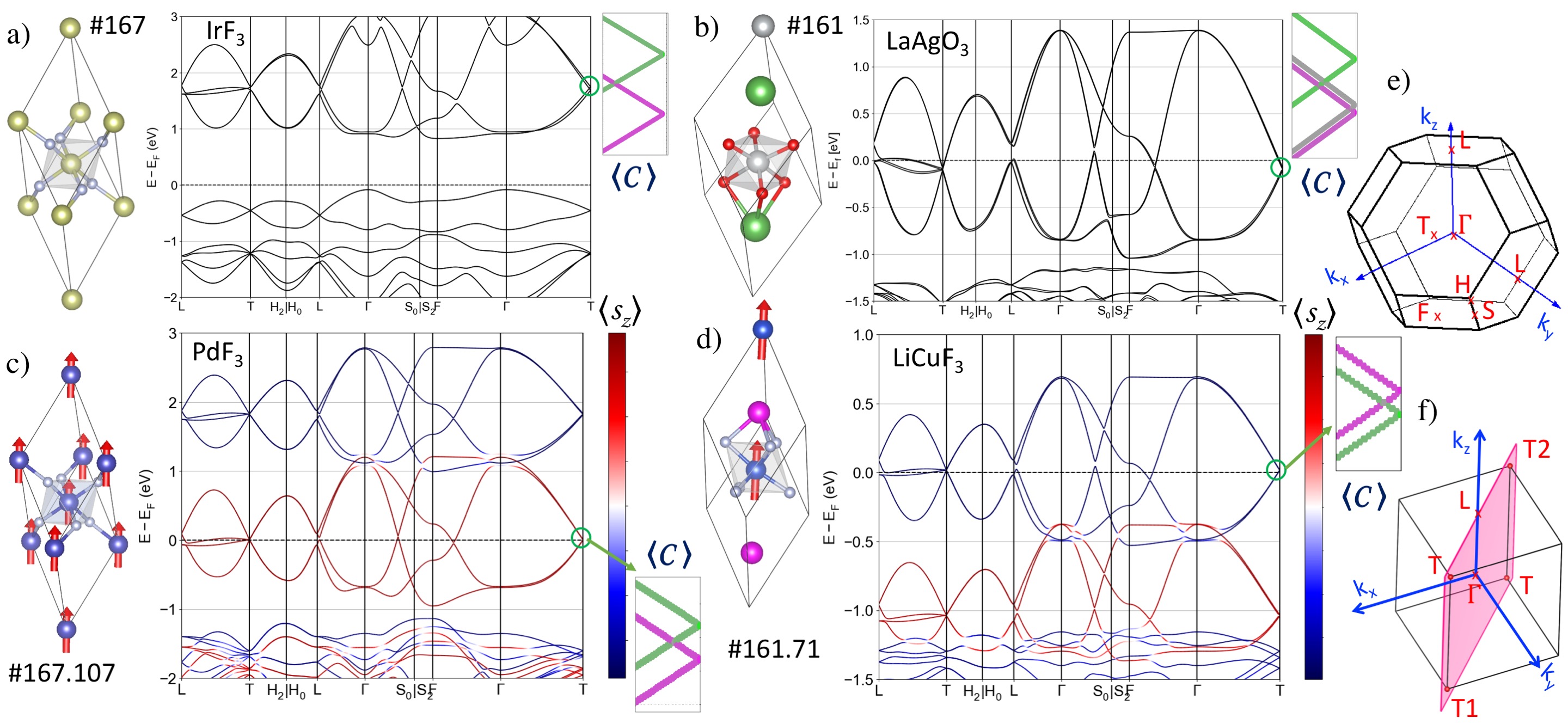}
	\caption{Rhombohedral crystal structure and electronic band structure calculated for (a) IrF$_3$ (space group \#167), (b) LaAgO$_3$ (space group \#161), (c) PdF$_3$ (magnetic space group \#167.107) and (d) LiCuF$_3$ (magnetic space group  \#161.71). Eigenvalues of the rotation $\langle \mathcal{C} \rangle$ and the spin-$z$  $\langle $S$_z\rangle$ operators projected on the electronic bands are shown. The zoom-in band structures near to the T point show the band crossings along the symmetry line $\Gamma$-T. (e) Brillouin zone of the rhombohedral structure with the labels of the high-symmetry points. (f) $(1\bar{1}0)$ plane on the reciprocal space of the rhombohedral structure.}
	 \label{Fig1}
\end{figure*} 

Rhombohedral crystals are structures that fit into rhombohedral Bravais lattices with space groups \#146, 148, 155, 160, 161, 166, or 167. 
 Materials with these crystal lattices can present a layered structure with a hexagonal order within each layer, which is common in
 topological insulators as Bi$_2$Te$_3$, Bi$_2$Se$_3$ and Sb$_2$Te$_3$ \cite{Zhang2009}. A broad kind of insulators and semiconductor materials as
 XF$_3$-type, ABF$_3$-type, ABO$_3$-type, and AB(PO$_4$)$_3$-type, crystallize in these rhombohedral phases with spaces 
groups \#161 (R3c) and \#167 (R$\bar{3}$c).
\\
In particular, IrF$_3$ presents the most stable phase in the space group \#167 with an indirect energy gap of $0.9$ eV and an electronic
 band structure as it is shown in Fig. \ref{Fig1}(a). The first conduction bands have a particular shape which is common in other materials
 with  space groups \#167 and \#161. This electronic band structure presents two close bands in the L-T high symmetry line of the Brillouin
 zone, see Fig. \ref{Fig1}(e), and depending on the material, the Fermi level can cross these close bands generating unique electronic and
 spin transport properties. This is the case of the trifluoride LaAgF$_3$, which crystallizes in the space group \#161 as it is shown in
 Fig.  \ref{Fig1}(b). The material LaAgF$_3$ has two LaAgF$_3$ molecules per unit cell and every molecule has five atoms. 
In this case, due to the loss of the inversion symmetry with respect to the space group \#167, the band structure presents unfolding.
\\
On the other hand, it was recently discovered that magnetic topological insulators may prefer the rhombohedral phases as is the case of
 the material MnBi$_2$Te$_4$ \cite{Deng895}. Among the materials that can crystallize in the magnetic rhombohedral-type structure we may find the 
 conductive transition-metal (TM) fluorides and oxides such as MnF$_3$, CoF$_3$, PdF$_3$, FeF$_3$, LiCuF$_3$, MnBO$_3$, LaMnO$_3$,
TiBO$_3$ and LaNiO$_3$ (See Table \ref{table four groups}). 
For the case of XF$_3$, the primitive unit cell contains two transition-metal atoms surrounded by six F atoms that form an
 octahedron as it is shown in Fig.  \ref{Fig1}(c) and (d). Recently, it was found that the TM atoms can exhibit a ferromagnetic 
order in the $z$ direction as it is shown in Fig. \ref{Fig1}(c), where the local magnetic moments are indicated by arrows. Fig. \ref{Fig1}
 also shows the band structure along high-symmetry lines including the spin-orbit coupling interaction. Similar behavior is observed for
 LiCuF$_3$ materials, which also present a ferromagnetic stable structure oriented parallel to the $z$ axis, as it is shown in Fig. \ref{Fig1}(d).
\\
\begin{figure*}
	\includegraphics[width=18.4cm]{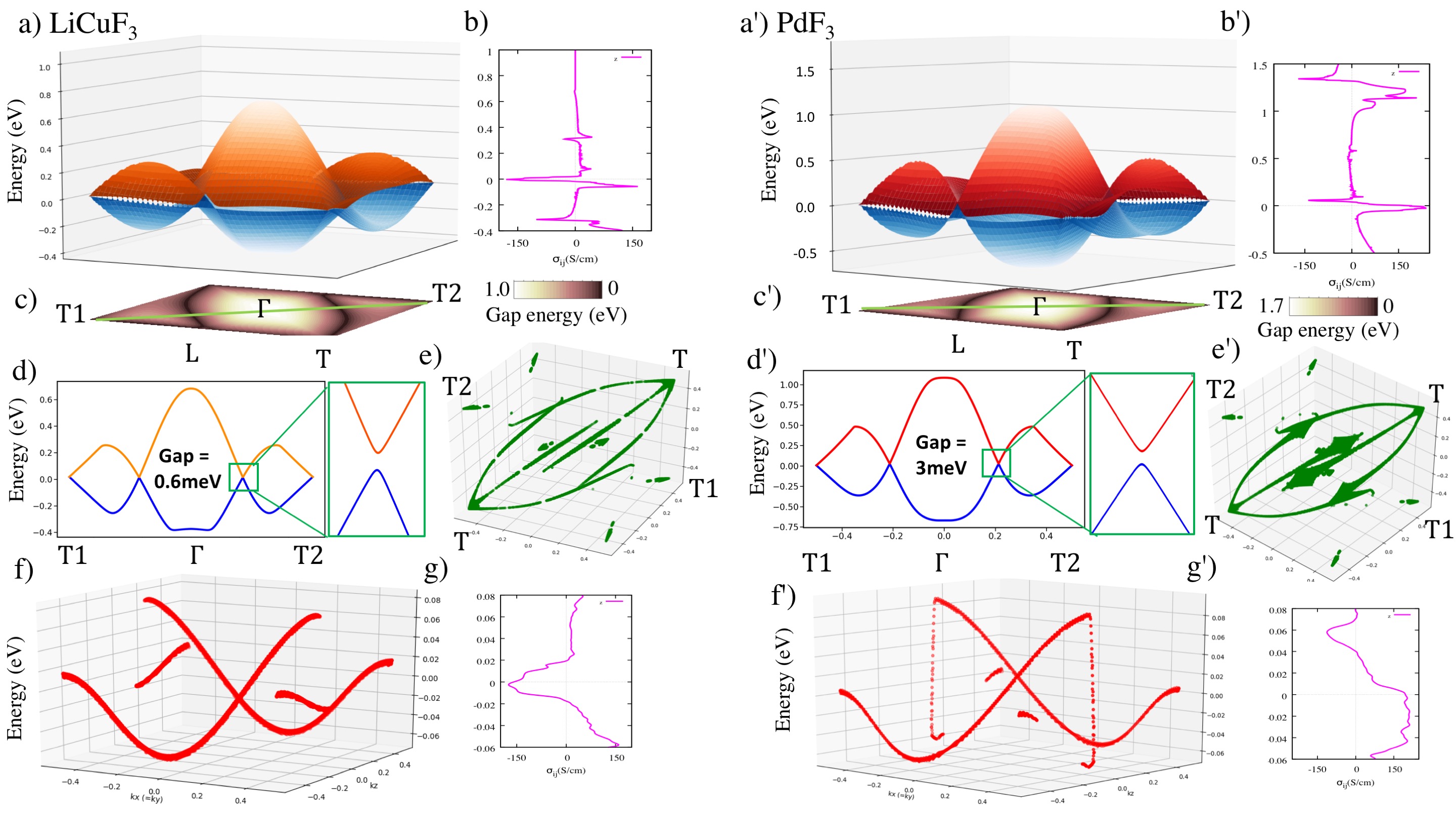}
	\caption{Quasi-nodal lines and Anomalous Hall conductance for the rhombohedric trifluorides LiCuF$_3$ and PdF$_3$.
		(a) and (a'), energy of the valence and conduction bands restricted to the fixed $\mathcal{G}$ plane $k_x=k_y$, (c) and (c') contour of the bandgap energy (difference between the conduction and the valence bands) and (f) and (f') quasi-nodal lines in the energy vs the plane $k_x(=k_y$) and $k_z$ space for the plane $(1\bar{1}0)$  shown in Fig.  \ref{Fig1} (f). (d) and (d') calculated band structure for the T1-$\Gamma$-T2 k-path, and zoom around the anticrossing point showing the bandgap energy. (e) and (e'), structure of the 3D quasi-nodal lines in momentum space in which it is noted the $C_3$ symmetry around the T-T axis. (b), (b'), (g) and (g') anomalous Hall conductivity as a function of the Fermi level for LiCuF$_3$ in the magnetic space group \#161.71 and for PdF$_3$ in the magnetic space group \#167.107.  The peaks of the AHC close to the Fermi level in diagrams (b) and (b') are located on the energy intervals where the quasi-nodal lines are defined. This fact can be appreciated in diagrams (f) and (g) for LiCuF$_3$ and (g') and (f') for PdF$_3$ where the location of the quasi-nodal lines on the energy level is compared to the AHC on the same energy.}
	\label{Fig2}
\end{figure*}  
\\
The projection of the  $z$-spin component is also shown in the band structure for the ferromagnetic cases. It can be noted a conductor
behavior for the up (down) spin orientation in PdF$_3$ (LiCuF$_3$), but as an insulator for the opposite orientation.  Consequently,
the spin band polarization of the conduction of electrons is 100\%, and therefore the longitudinal and transverse currents will be full spin-polarized.  For PdF$_3$ and LiCuF$_3$ is obtained an opposite spin bandgap of around 2.14 eV and 4.10 eV, respectively.
These results are in agreement with previous theoretical data \cite{PhysRevLett.119.016403,PhysRevLett.124.016402,WANG2019554}. 
It is worth remarking that the spin-polarized band structure at the Fermi level shows a similar shape as in the nonmagnetic case.  
In these cases, the Fermi level crosses the nearby bands in the L-T path, which generates band energies and bandgap contours at the plane $k_x=k_y$ as is shown in Fig. \ref{Fig2}(a) and (a'), and  Fig. \ref{Fig2}(c) and (c').  It is also observed a band crossing between valence and conduction bands close to the T point. This special feature of the band structure can induce a large anomalous Hall conductivity (AHC) and spin Hall conductivity (SHC) at the Fermi level for the ferromagnetic and non-magnetic cases, respectively.
\\
In the presence of SOC, the band structure and the symmetry group of the materials depend on the direction of the total magnetization. In the PdF$_3$ and LiCuF$_3$ cases, the lowest energy was found for the magnetization in the $z$ direction.  For the XF$_3$ case with $\mathbf{m}\parallel z$, the \#167 systems present the 167.107 magnetic space group with symmetries generated by $\mathcal{I},\mathcal{C}$  and the anti-unitary symmetry $\mathbb{T} \mathcal{G}$.  In the ABF$_3$ case with $\mathbf{m}\parallel z$, the \#161 systems present the 161.71 magnetic space group with symmetries generated by $\mathcal{C}$ and the anti-unitary symmetry 
$\mathbb{T}\mathcal{G}$.  
\\
In the cases of XF$_3$ and ABF$_3$ with magnetic space groups 167.107 and 161.71, the multiple symmetries prohibits the anomalous Hall effect (AHE) in $x$ and $y$ components. The only component which is not constrained is $b_{z}(\textbf{k}) \neq 0$, so the AHE should be observed with $\sigma_{xy} \neq 0$ as it is shown in Fig. \ref{Fig2}(b) and (b').  These graphs show the variation of AHC for PdF$_3$ and LiCuF$_3$ with respect to the position of the Fermi energy. It is worth noting that the peaks of the AHC around the Fermi energy for $\sigma_{xy}$ are as high as 180 S/cm. As it was pointed out before, close to the Fermi energy we have a full spin-polarized charge current and therefore we can consider the $z$-spin component as a good quantum number \cite{spin-polarization}. So the Hall current carriers are one spin-polarized electron and the SHC can be calculated from the AHC by a factor of $2\hbar/e$, i.e. $\sim$$90 (\hbar/e)(S/cm)$. These results were corroborated with the direct calculation of the SHC tensor.
\\
In addition, it is found that the main contribution to the AHC at the Fermi level is due to the existence of the quasi-nodal lines close to the Fermi level.  The main contribution to the AHC at top and bottom values of the Fermi energy is the electronic states generated by the quasi-nodal lines as it is shown in Fig. \ref{Fig2}(g) and (g'). In Fig. \ref{Fig2}(f) and (f') it is shown the energy distribution of the quasi-nodal lines as a function of $k_x(=k_y)$ and $k_z$; these energies match the strong signal of the AHC in the energy window. The reciprocal space distribution of the quasi-nodal lines is shown in Fig.  \ref{Fig2}(e) and (e'), which shows clearly the six-fold multiplicity as indicated in the Table \ref{table four groups} and $C_3$ rotation symmetry around the T-T $k$-path. This corresponds to the main axis of  $C_3$ symmetry in real space that characterizes the rhombohedral lattice structures. Finally, it is noted the anticrossing band gap between the energy bands in Fig.  \ref{Fig2}(d) and (d') along the quasi-nodal line on the $k_x(=k_y)$ and $k_z$ plane for the LiCuF$_3$ (0.6 meV) and PdF$_3$ (3.0meV).
\\

\section{Computational method}

We have carried out $ab$-$initio$ calculations within the density-functional theory (DFT) framework to study the formation of quasi-nodal lines in magnetic and nonmagnetic rhombohedral materials.  Exchange and correlation effects were treated with generalized gradient approximation (GGA) \cite{Perdew1996}, as implemented in the vienna \textit{ab-initio} simulation package (VASP) \cite{vasp}. The GGA+U method were employed for the PdF$_3$ material as presented in a recent report \cite{PhysRevLett.124.016402}. Spin orbit coupling (SOC) were included self-consistently in all the calculation.  The electron wave function was expanded in plane waves up to a cut-off energy of 520 eV . A $k$-mesh of 0.02 (2$\pi$/\AA) $k$-space resolution was used to sample the Brillouin zone.   DFT calculations of rhombohedral materials were performed using the refined lattice constants from the Materials project database \cite{materialsproject}. In addition, we calculated the symmetry eigenvalues of the wavefunctions at the Brillouin zone using the irrep code \cite{irrep}.

In order to evaluate the electronic transport properties, we have used  an effective tight-binding Hamiltonian constructed in the maximally localized Wannier basis \cite{wannier90} as a post-processing step of the DFT calculations.  The intrinsic anomalous Hall conductivity (AHC) components were calculated by integrating the Berry curvature on a dense $230^3$ $k$-mesh of the Brillouin zone \cite{wb}. Within this model, the AHC can be written as:
\begin{equation}
\sigma_{xy}=-\dfrac{e^2}{\hbar} \sum_{n}\int_{BZ}\dfrac{dk^3}{(2\pi)^3}f_n(k)\Omega^z_n(k)
\end{equation}
Where $f_n(k)$ is the Fermi-Dirac distribution and the Berry curvature $\Omega^z_n(k)$ for the $n$th band can be calculated using the Kubo formula:
\begin{equation}
\Omega^z_n(k)= -2{\hbar}^2 \text{Im}\sum_{m\neq n}%
\frac{\left\langle n,k\vert \widehat v_x \vert m,k \rangle\langle m,k\vert \widehat v_y \vert n,k  \right\rangle}{\left(\epsilon_{n,k}-\epsilon_{m,k}\right)^{2}} 	
\end{equation}
Where $\vert n,k\rangle$ are the Bloch functions of a single band $n$, $k$ is the Bloch wave vector, $\epsilon_{n,k}$ is the band energy and $\hat{v}_i$ is the velocity operator in the $i$ direction.

\section{Discussion and conclusions}

In this work we have put forward the concept of quasi-nodal lines. These are lines on the reciprocal space where the 
energy gap is very small. They appear due to the hybridization of the nodal lines that glide reflection symmetries induce. The hybridization
may be due to the inclusion of inversion symmetry, or due to magnetization which breaks the glide reflection symmetry. In both cases, if the
gap generated by the hybridization is small, the quasi-nodal lines are present. In the family of rhombohedral trifluorides and trioxides, both
ferromagnetic and non-magnetic, we have shown the existence of these quasi-nodal lines. Whenever the quasi-nodal lines
are close to the Fermi level, interesting electronic transport properties are induced. This is the case of the ferromagnetic phase of
PdF$_3$ and of LiCuF$_3$ where the presence of the quasi-nodal line on the Fermi level induces a large signal in the anomalous Hall conductance.

Quasi-nodal lines are not topologically protected. By this we mean that the energy gap could be adiabatically enlarged. Nevertheless, 
in practice the energy gap depends on the material structure and on the intensity of the SOC. Herein, by using  \textit{ab-initio} calculations, we have shown that the energy gap remains small for the materials mentioned above and therefore they could be detected experimentally.  For the  ferromagnetic phases, we have demonstrated a direct correspondence between the AHE (or SHE) signal and the existence of the quasi-nodal lines. 
%the large signal of the anomalous Hall conductance evidences the existence of the quasi-nodal lines.

It is interesting to determine other families of materials on which quasi-nodal lines are also present. These families of materials
may not crystallize with symmetry groups on which the nodal lines deem to exist, ie. the symmetry groups may not be on
the list of symmetry groups with nodal lines. Nevertheless, the presence of quasi-nodal lines induce interesting electronic
transport properties and therefore it is worth their future experimental research.

Quasi-nodal lines in ferromagnetic phases are due to the hybridization of nodal lines because the glide reflection symmetry is broken. 
It is interesting to note that the combinatorial structure of the bands is similar to the one of the material prior
to the magnetization. This would mean that the magnetization, even though it breaks symmetries, it still remembers some of the
information of the non-magnetic group. The reconstruction of quasi-nodal lines from the ferromagnetic symmetries is not straightforward; 
while the knowledge of the non-magnetic symmetries permits to infer the possible existence of these quasi-nodal lines.

%*** Acknowledgement ***
\section{Acknowledgements}
The first author gratefully acknowledges the computing time granted on the supercomputer Mogon at Johannes Gutenberg University Mainz (hpc.uni-mainz.de). The second author thanks the German Service of Academic Exchange (DAAD) for its continuous support.  The first and the third authors thank the continuous support of the Alexander Von Humboldt Foundation, Germany.

%*** Bibliography ***
%\newpage

%\bibliographystyle{abbrv}
\bibliographystyle{apsrev}
\bibliography{bibliography}
%\bibliography{topological.bib}

\end{document}